\documentclass{article}

\usepackage{arxiv}

\usepackage[utf8]{inputenc} % allow utf-8 input
\usepackage[T1]{fontenc}    % use 8-bit T1 fonts
\usepackage{hyperref}       % hyperlinks
\usepackage{url}            % simple URL typesetting
\usepackage{booktabs}       % professional-quality tables
\usepackage{amsfonts}       % blackboard math symbols
\usepackage{nicefrac}       % compact symbols for 1/2, etc.
\usepackage{microtype}      % microtypography
\usepackage{lipsum}		% Can be removed after putting your text content
\usepackage{graphicx}
\usepackage{natbib}
\usepackage{doi}

\usepackage[table, dvipsnames]{xcolor}
\usepackage{multirow}

\usepackage{amsmath}
\usepackage{amssymb}
\usepackage{comment}
\usepackage{subcaption}
\usepackage{caption}
\usepackage[utf8]{inputenc}
\usepackage{soulutf8}
\usepackage{pifont}
\usepackage{amsmath}
\usepackage{amssymb}
\usepackage{bm}
\usepackage{color,soul}
\usepackage[dvipsnames]{xcolor}

\usepackage{xcolor}
\usepackage{hyperref}
\hypersetup{
    colorlinks,
    linkcolor={red!80!black},
    citecolor={blue!50!black},
    urlcolor={blue!80!black}
}

\newcommand{\mytilde}{\raise.17ex\hbox{$\scriptstyle\mathtt{\sim}$}}

\def\bb{\mathbf{b}}

\def\xx{\mathbf{x}}

\def\Re{\mathbb{R}}

\newcommand{\result}[2]{\ensuremath{#1}\scriptsize{$\pm$\ensuremath{#2}}}

\title{FeDETR: a Federated Approach for Stenosis Detection in Coronary Angiography}

\author{
    \href{https://orcid.org/0000-0002-1171-5672}{\includegraphics[scale=0.06]{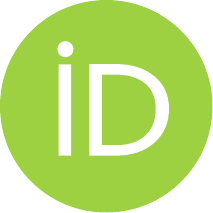}\hspace{1mm}
    Raffaele~Mineo}\thanks{Equal contribution} \\
    Department of Engineering\\
    University Campus Bio-Medico of Rome\\
    Rome, Italy\\
	\texttt{raffaele.mineo@unicampus.it} \\
	\And
    \href{https://orcid.org/0009-0001-8539-6299}{\includegraphics[scale=0.06]{orcid.pdf}\hspace{1mm}
    Amelia~Sorrenti}$^{*}$ \\
    Department of Engineering\\
    University Campus Bio-Medico of Rome\\
    Rome, Italy\\
	\texttt{amelia.sorrenti@unicampus.it} \\
	\And
	\href{https://orcid.org/0000-0002-6122-4249}{\includegraphics[scale=0.06]{orcid.pdf}\hspace{1mm}Federica~Proietto Salanitri} \\
	Department of Engineering\\
	University of Catania\\
	Catania, Italy\\
	\texttt{federica.proiettosalanitri@unict.it} \\
}

\date{}

\renewcommand{\shorttitle}{FeDETR: a Federated Approach for Stenosis Detection in Coronary Angiography}

\hypersetup{
pdftitle={FeDETR: a Federated Approach for Stenosis Detection in Coronary Angiography},
pdfsubject={},
pdfauthor={Raffale~Mineo, Amelia~Sorrenti, Federica~Proietto Salanitri},
pdfkeywords={federated learning, coronary angiography, medical imaging analysis, stenosis detection},
}

\begin{document}
\maketitle

\begin{abstract}
	Assessing the severity of stenoses in coronary angiography is critical to the patient's health, as coronary stenosis is an underlying factor in heart failure. Current practice for grading coronary lesions, i.e. \emph{fractional flow reserve} (FFR) or \emph{instantaneous wave-free ratio} (iFR), suffers from several drawbacks, including time, cost and invasiveness, alongside potential interobserver variability.
In this context, some deep learning methods have emerged to assist cardiologists in automating the estimation of FFR/iFR values. Despite the effectiveness of these methods, their reliance on large datasets is challenging due to the distributed nature of sensitive medical data.
Federated learning addresses this challenge by aggregating knowledge from multiple nodes to improve model generalization, while preserving data privacy.
We propose the first federated detection transformer approach, \emph{FeDETR}, to assess stenosis severity in angiography videos based on FFR/iFR values estimation. In our approach, each node trains a detection transformer (DETR) on its local dataset, with the central server federating the backbone part of the network.
The proposed method is trained and evaluated on a dataset collected from five hospitals, consisting of 1001 angiographic examinations, and its performance is compared with state-of-the-art federated learning methods.

\end{abstract}

\keywords{federated learning \and coronary angiography \and medical imaging analysis \and stenosis detection}

\section{Introduction}
One of the main causes of heart failure is \emph{coronary stenosis}, which occurs when the blood vessels narrow and prevent the normal pumping of blood~\cite{neumann2018guidelines,knuuti20202019}. Depending on the severity of the stenosis, the first step in the evaluation of a coronary angiography is to decide whether medical or surgical treatment is required.

Invasive assessment of coronary physiology, using either \emph{fractional flow reserve} (FFR) or \emph{instantaneous wave-free ratio} (iFR), is an established guideline method for grading coronary lesions~\cite{neumann2018guidelines,knuuti20202019}. However, these methods have certain limitations, such as time consumption, high costs, and potential complications due to their invasive nature. In addition, inter-observer variability in clinical decision making may occur depending on the expertise of the examiner in locating major stenosis~\cite{singh2003intra}.

Despite these drawbacks, quantification of stenoses severity using FFR and iFR has gained popularity in guiding revascularization strategies for multivessel disease. Studies suggest that FFR values below 0.80 and iFR values below 0.89 are indicative of hemodynamically-significant stenosis~\cite{neumann2018guidelines,tonino2009fractional,de2012fractional,davies2017use}, while patients with values above these thresholds may not benefit significantly from revascularization compared with optimal treatment alone. Therefore, automating the estimation of FFR or iFR values would assist clinicians in making accurate decisions and reducing the number of treatments patients need to undergo.

Over the last decade, the field of medical image analysis has seen the emergence of several deep learning methods to assist cardiologists in a variety of tasks (e.g., cardiovascular imaging analysis and risk assessment~\cite{litjens2019state,motwani2017machine}, and automated/semi-automated quantification of artery stenosis in coronary angiography~\cite{zhang2019direct,zhang2020direct,xue2018full}). In the context of stenoses quantification, the most promising approaches~\cite{zhang2020direct,zhang2019direct} take advantage of multiple angiography views in conjunction with the concept of \emph{key frame}, that is the highest quality video frame characterized by complete penetration of the contrast agent. %, resulting in clearly contrasted vessel borders. 
Although these techniques have been proven to be effective, they require datasets consisting of a large number of patient examinations, besides the need for physicians to manually identify key frames.

However, due to the sensitive nature of medical data, collecting and sharing large amounts of patient data distributed across multiple healthcare institutions and research centers is challenging.
Federated Learning~\cite{mcmahan2017communication,li2020federated,li2021fedbn} (FL) supports the medical domain by offering approaches that leverage the potential of distributed healthcare data and ensure that this data remains local. FL aggregates knowledge from multiple nodes, even when each individual dataset is small. This collective learning approach improves the generalization and performance of the model by taking advantage of the diversity of data from different sources.

To overcome the above-mentioned limitations, we propose an approach named \textbf{FeDETR} for the assessment of stenosis severity from angiography videos by taking advantage of the federated setting. In our training procedure, each node in the centralized federation trains a detection transformer (DETR)~\cite{carion2020end} on its own small heterogeneous dataset, while the central server federates the backbone network.
The proposed method detects hemodynamically significant stenosis in key frames using both direct and indirect estimation of FFR or iFR values. 
To train and evaluate the performance of our approach, the dataset used has been collected in five prominent hospitals and includes a total of 1001 angiographic examinations.
We assess the validity of the proposed approach by comparing it to state-of-the-art federated methods (i.e., FedAvg~\cite{mcmahan2017communication} and FedBN~\cite{li2021fedbn}).

\section{Related Work}
Federated Learning~\cite{mcmahan2017communication} aims to develop models that can be used to exploit the collective knowledge of multiple nodes, while preserving data privacy and security. 
In a classic FL setting, multiple client nodes fine-tune a base model received from a central server using local data and send back local model updates. The central server then aggregates these updates into a global model, which is iteratively sent back to the nodes until convergence is reached. This aggregation can be performed using several methods.

FedAvg~\cite{mcmahan2017communication} averages the model updates received from all participating devices to create the updated global model. FedProx~\cite{li2020federated} extends such an approach by introducing a proximal term into the optimization objective, maintaining similarity between the global model and local models during aggregation. Federated Median~\cite{pillutla2022robust}, as an alternative to FedAvg~\cite{mcmahan2017communication}, performs the aggregation by taking the median value of the model updates received from participating nodes, reducing the impact of outliers or noisy updates. In FedBN~\cite{li2021fedbn}, the central node aggregates model parameters while keeping batch normalization layers private.

In the context of FL, most of the methods used to perform object detection involve the use of YOLOv3~\cite{redmon2018yolov3} and Faster R-CNN~\cite{ren2015faster}, as proposed in~\cite{jallepalli2021federated,luo2019real}. FedVision~\cite{liu2020fedvision} is a federated proprietary algorithm based on YOLOv3~\cite{redmon2018yolov3} that enables end-to-end joint training of object detection models with locally stored datasets from multiple clients. Similarly, Yu and Liu~\cite{yu2019federated} applied the FedAvg~\cite{mcmahan2017communication} algorithm to the Single Shot MultiBox Detector (SSD)~\cite{liu2016ssd}, enhancing it with the Abnormal Weights Clip in order to reduce the influence of non-IID data.

Similarly, existing classical methods are used and adapted for federated image segmentation tasks. Tedeschini et al.~\cite{tedeschini2022decentralized} and Yi et al.~\cite{yi2020net} proposed two different modified versions of the U-NET~\cite{ronneberger2015u} architecture to perform brain tumor segmentation.
Li et al.~\cite{li2019privacy} compared several weight sharing strategies to mitigate the effects of data imbalance across different hospitals and to reduce the potential risk of training data being reverse-engineered from the model parameters. FedDis~\cite{bercea2021feddis} is a federated method for segmentation that disentangles the parameter space into shape and appearance, and shares only the shape parameter across clients, mitigating domain shifts among individual clients.

Since the medical field is characterized by distributed and heterogeneous data, it can be used as test bench for federated learning methods~\cite{feki2021federated,dayan2021federated,li2019privacy}. FL provides a privacy-preserving approach that enables healthcare institutions to collaborate and build robust and generalized models by training on diverse data from different patient populations.

Over the past few years, several deep learning methods have been employed to detect and quantify the severity of stenosis from imaging data. These methods can be based on \textit{2D} or \textit{3D approaches}. The former analyzes individual frames previously extracted from an angiography video and then performs the final prediction. The latter, which has been less researched, extracts the spatio-temporal features directly from the whole video~\cite{zhang2019direct,zhang2020direct,xue2018full}.

The vast majority of the 2D approaches perform stenosis classification by grading into two or three severity levels or by thresholding FFR/iFR values in order to detect hemodynamically significant stenosis.  These methods typically rely on CNN architectures~\cite{moon2021automatic,antczak2018stenosis} or a combination of convolutional and recurrent networks~\cite{cong2019automated,ma2017fast,ovalle2022hybrid} to automatically identify \emph{key frames} and then employ a stenosis classification module for final classification. 

Alternatively, some approaches locate stenosis by focusing only on blood vessels in the key frame, i.e., by highlighting them in a pre-processing segmentation step~\cite{wu2020automatic,au2018automated} or by analyzing their shape and visual appearance~\cite{zhao2021automatic,zhao2021new,wan2018automated}. Zhao et al.~\cite{zhao2021automatic} perform automatic vessel segmentation, followed by keypoint extraction and classification to identify segments with a high likelihood of stenosis.

In addition, some 2D methods apply interpretability techniques on frame-based stenosis classification models to generate activation maps that help guide the stenosis detection process~\cite{moon2021automatic,cong2019automated}.

Differently from the aforementioned methods, our approach combines detection transformers (DETR)~\cite{carion2020end} with federated learning to assess stenosis severity from angiographic key frames.

\section{Method}
\begin{figure}[!b]
    \includegraphics[width=\textwidth]{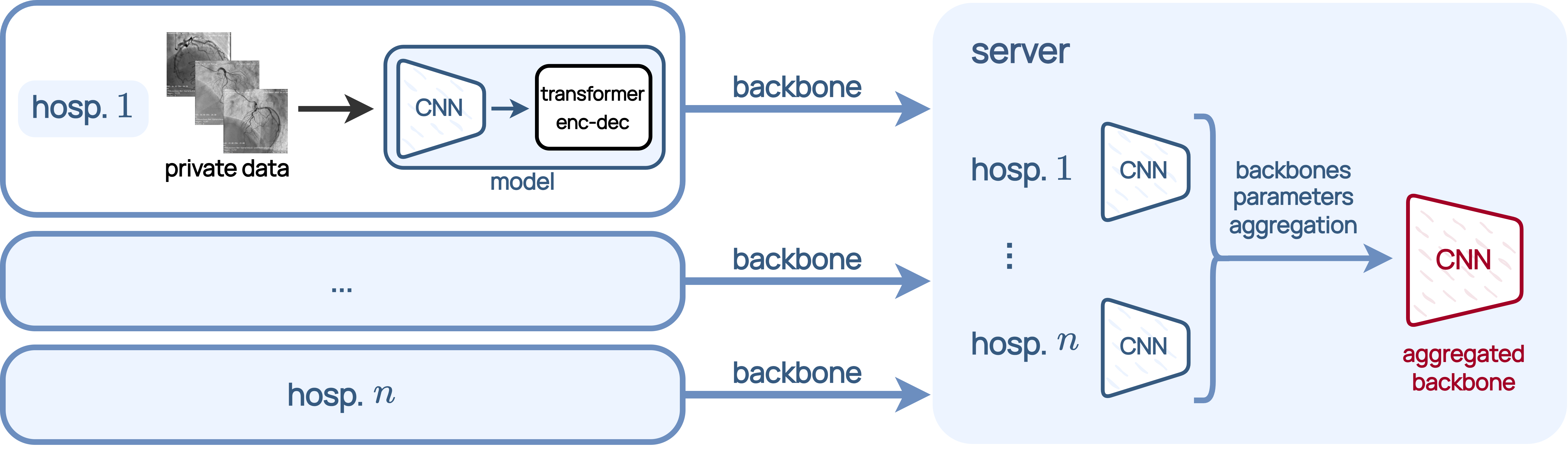}
    \caption{Each node uses its own dataset, consisting of key frames extracted from coronary angiographies videos, to train a detection transformer (DETR). At the beginning of each round, the central server sends the aggregated backbone to each node, which uses it to extract features from the key frames. After each round, each node sends its locally trained backbone back to the central server, which aggregates them all.} 
    \label{fig1}
\end{figure}

\begin{figure}[ht!]
    \includegraphics[width=\textwidth]{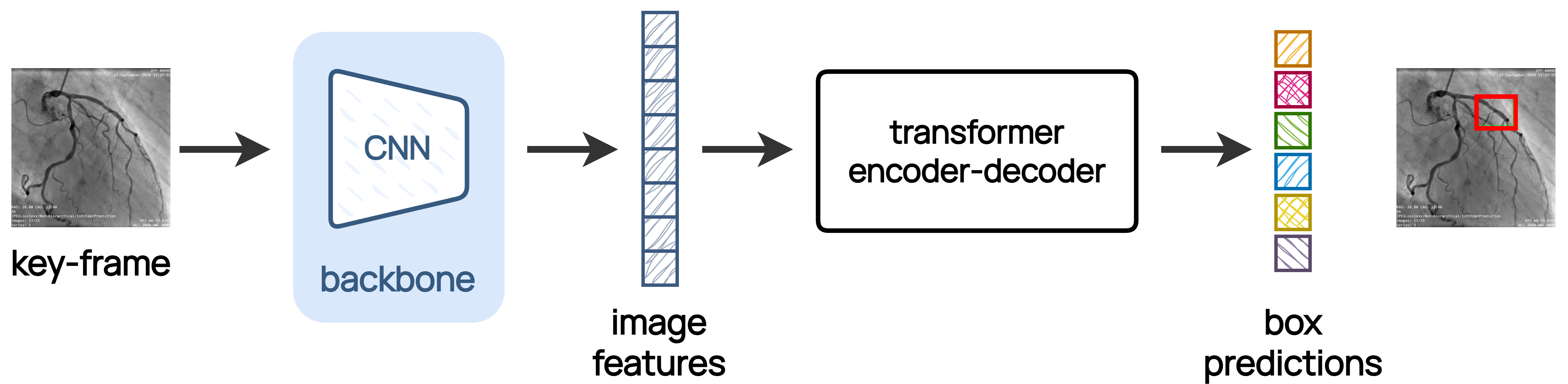}
    \caption{Given a key frame, the CNN backbone learns a 2D representation, which is then flattened by the model. Before passing these learned features to the transformer encoder, the model also extends them with positional encoding. The transformer decoder outputs some learned positional embeddings and feeds them into a feed-forward network that predicts classes and their corresponding bounding boxes.} 
    \label{fig2}
\end{figure}

An overview of the proposed approach is shown in Fig.~\ref{fig1}.
In this scenario, a federation consists of a set of $n$ peer nodes managed by a central server. Each node owns a private dataset consisting of \emph{key frames} extracted from coronary angiography videos. This dataset is used to train a detection transformer (DETR)~\cite{carion2020end}, as shown in Fig.~\ref{fig2}.

During each round of centralized training, each node receives the backbone of the model from the server. %The input backbone is used to extract features from the private images, which are then fed into the transformer part of the model.
The backbone is used to extract features from the images in the private dataset. After being flattened by the model and enhanced with position encoding, these features are fed into the encoder-decoder transformer.
Each of the $N$ output embeddings of the transformer decoder is then passed to a shared feed-forward network, which independently decodes it into bounding boxes and class labels, resulting in $N$ final predictions.

At the end of each round, which involves multiple training iterations, the locally trained backbone from each device is sent back to the central server. The server performs the aggregation of the received backbones, and the entire process is repeated for the next round of training. This iterative process ensures collaboration between nodes while preserving data privacy and security.

\subsection{Problem formulation}

The proposed method operates in a federated setting, where $K$ nodes jointly train for $R$ rounds. At the end of each round, which consists of $T$ local epochs, a communication with a central server is performed.

Each node $i$ trains a local model $M_i$ on its own private dataset $\mathcal{D}_i = \{(\xx_1, y_1, $ $ \bb_1), (\mathbf{x}_2, y_2, \bb_2), \ldots,(\mathbf{x}_n, y_n, \bb_n)\}$, where $\mathbf{x}_j\in \mathcal{X}$ represents a sample image within the dataset, $y_j\in \mathcal{Y}$ represents the target object class (low-severity or mild/high-severity stenosis), and $\bb_j \in \Re^4$ represents bound box coordinates.
%Hence, the round $r_i$ aims at optimizing the model $M_i$ on dataset $D_i$ residing on node $k_i$ and that cannot be shared to other nodes.
Each local model $M_i=\langle g, \boldsymbol{\theta}_i\rangle$ is parameterized by $\boldsymbol{\theta}_i$, with $g: \mathcal{X} \rightarrow \mathcal{Y}$. Specifically, in the proposed approach, the $K$ nodes train a local DETR model~\cite{carion2020end} whose architecture is illustrated in Fig.~\ref{fig2}.
Since the model consists of two parts, the model parameters $\boldsymbol{\theta}_i$ can be written as the union of the backbone and the encoder-decoder transformer parameters, $\boldsymbol{\phi}_i$ and $\boldsymbol{\omega}_i$ respectively.

Hence, at training time, each node $i$ in the federation, locally optimizes a finite sum objective in the form:

\begin{equation}
\label{eq:loss}
    \min_{\boldsymbol{\theta}_i} f(\boldsymbol{\theta}_i)~~~\text{with}~~~f(\boldsymbol{\theta}_i) \equiv f(\boldsymbol{\phi}_i, \boldsymbol{\omega}_i) \text{,}
\end{equation}
defining $f(\boldsymbol{\phi}_i, \boldsymbol{\omega}_i)$ as follows:

\begin{equation}
    f(\boldsymbol{\phi}_i, \boldsymbol{\omega}_i) = \frac{1}{n} \sum_{j=1}^{n} f_j(\boldsymbol{\phi}_i, \boldsymbol{\omega}_i) \text{,}
\end{equation}
where $f_j(\boldsymbol{\phi}_i, \boldsymbol{\omega}_i)=\ell(g(\mathbf{x}_j), y_j, \bb_j; \boldsymbol{\phi}_i, \boldsymbol{\omega}_i)$ denotes an object detection loss for $(\mathbf{x}_j, y_j)$ using the model parameters $\boldsymbol{\theta}_i$.

In contrast to current federated approaches, which are mostly based on the averaging of all model parameters, our method only federates the backbone part of the model. Indeed, at the end of each training round, the nodes in the federation apply the FedAvg algorithm~\cite{mcmahan2017communication} on the subset of parameters $\left\{ \boldsymbol{\phi}_1, \dots, \boldsymbol{\phi}_K \right\}$, thus receiving an updated version of the backbone and resuming local training.

\section{Experimental Evaluation}

\subsection{Dataset}
The proposed method was trained and evaluated using a private dataset comprising of 1001 coronary angiographies from patients with chronic coronary syndrome (CCS) and acute coronary syndrome (ACS). The dataset was collected in five large hospitals between 2020 and 2022, and included 737 male and 264 female patients with an average age of 66.8 $\pm$ 9.26 years.

The coronary angiographies and the physiological measurements were carried out in accordance with the standard clinical practice. Then, two experienced cardiologists annotated the key frames in the angiography data.
Each angiography was evaluated using invasive physiological assessment with iFR, FFR, or both. Specifically, FFR values were available for 613 patients (61.2\%), iFR values for 667 patients (66.6\%), and a subset of 279 patients (27.8\%) had both FFR and iFR data. Cardiologists identified the major stenosis in each exam and labeled it as hemodynamically significant if the FFR was less than 0.80~\cite{neumann2018guidelines,tonino2009fractional,de2012fractional} or if the iFR was less than 0.89~\cite{neumann2018guidelines,davies2017use}. Consequently, the 40.5\% of patients has been labeled as positive, resulting in a significant class imbalance.

\subsection{Training procedure}
% \lipsum[2-3]

Each node trains a DETR model~\cite{carion2020end}, consisting of a ResNet-50~\cite{he2016deep} backbone and an encoder-decoder transformer. The model was pre-trained on the COCO 2017 dataset~\cite{lin2014microsoft}. %The feed-forward network classifies each bounding box into three classes, including ``no object'', ``low risk stenosis'', and ``mild or high risk stenosis''.
The DETR models were trained by minimizing the Hungarian loss~\cite{carion2020end} using the AdamW~\cite{loshchilov2017decoupled} optimizer with initial learning rates set to 10$^{-4}$ and 10$^{-5}$ for the transformer and the backbone, respectively, and weight decay set to 10$^{-4}$. The learning rate is then updated using a scheduler with the multiplicative factor $\gamma$ set to 0.1. Gradient clipping was also applied with a maximum gradient norm of 0.1.

Both the encoder and the decoder consist of six layers characterized by a standard architecture. Their self- and cross-attention modules have eight heads, each receiving input embeddings of size 256. In addition, the feed-forward network (FFN) of these modules has a dimensionality of 2048. An additive dropout of 0.1 is applied after each multi-head attention and FFN and before performing the layer normalization. The output of the encoder is supplemented with a sinusoidal positional encoding before being passed to the decoder. The number of object queries of the transformer decoder is 20.
% The number of object queries that are passed as input to the transformer decoder is set to 20.

\begin{figure}
    \centering
    \begin{subfigure}[b]{.56\textwidth}
    \centering
    \includegraphics[width=.3\textwidth]{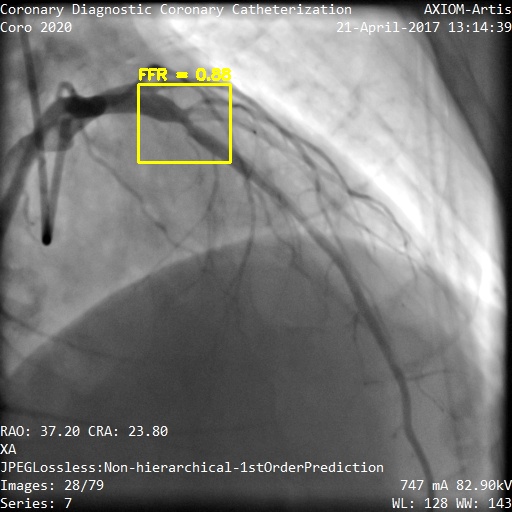}
    \includegraphics[width=.3\textwidth]{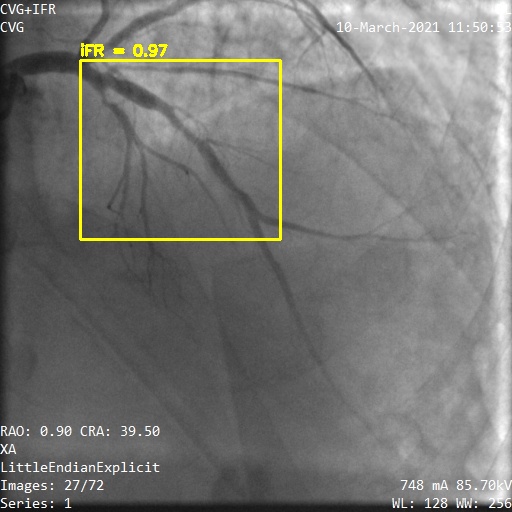}
    \includegraphics[width=.3\textwidth]{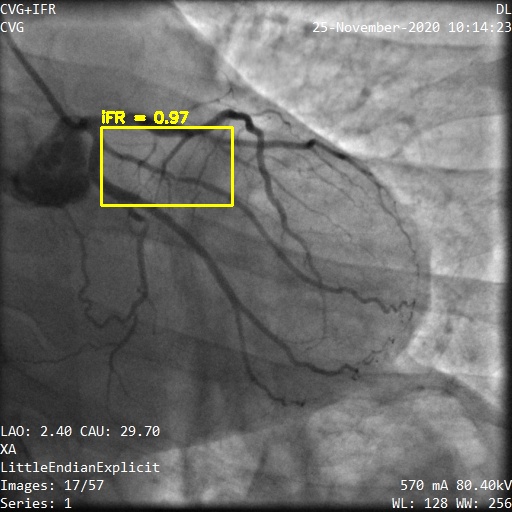}\\
    \includegraphics[width=.3\textwidth]{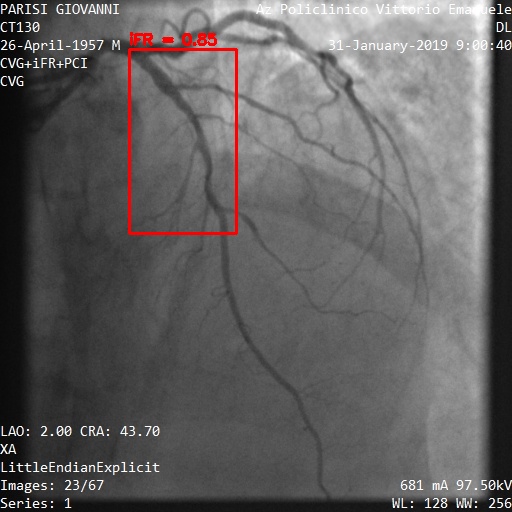}
    \includegraphics[width=.3\textwidth]{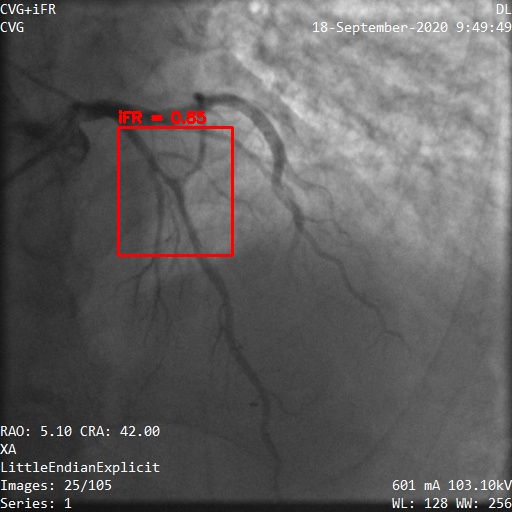}
    \includegraphics[width=.3\textwidth]{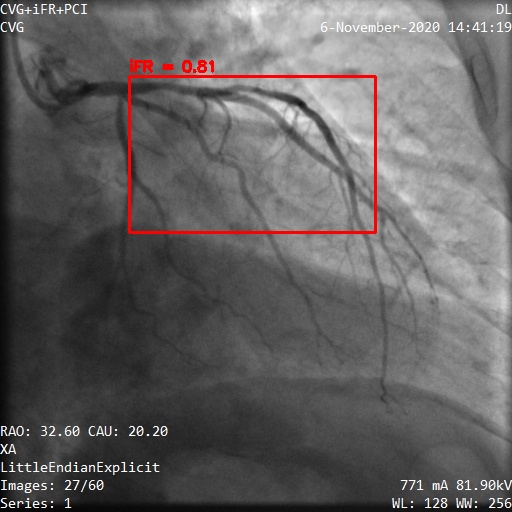}
    \caption{}
    \label{fig:stenosis}
    \end{subfigure}
    \begin{subfigure}[b]{.40\textwidth}
    \centering
    \includegraphics[width=\textwidth]{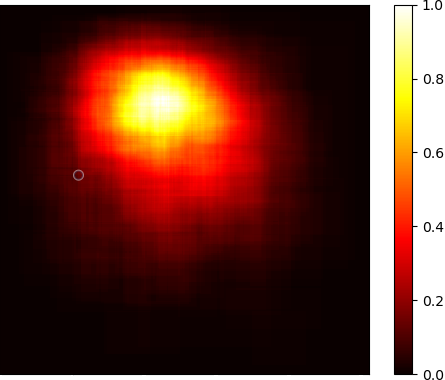}%\quad    
    \caption{}
    \label{fig:heatmap}
    \end{subfigure}
    \caption{%Figure~\ref{fig:stenosis} shows some samples of \emph{low severity stenoses} (on the top row) and \emph{mild or high severity stenoses} (on the bottom row). Figure~\ref{fig:heatmap} shows the heatmap of all the bounding box positions.
    Figure~\ref{fig:stenosis} shows representative examples of \emph{low severity stenoses} (illustrated in the top row) and \emph{mild or high severity stenoses} (depicted in the bottom row). Figure~\ref{fig:heatmap} presents an averaged heatmap, which is derived from the collective bounding box positions across all samples.}
    \label{fig3} 
\end{figure}

Data augmentation is performed by consecutively and randomly applying horizontal and vertical flips, and 90-degree rotations. Scale augmentation is then applied to the images, resizing them so that the shortest side is 512 pixels and the longest side is 1333 pixels or less.
Furthermore, zero padding is also added to the images, making them rectangular in order to shift the detection area. Since bounding boxes are usually positioned in the upper left corner or centrally, as demonstrated by the heatmap in Fig.~\ref{fig:heatmap}, the additional padding should prevent the model from learning a location bias.
\begin{comment}
\begin{figure}[!t]
    \centering
    \includegraphics[width=\textwidth]{figures/heatmap_stenoses.png}
    \caption{(a) The figure shows some samples of \emph{low severity stenoses} (i.e., the yellow bounding boxes) and \emph{mild or high severity stenoses} (i.e., the red bounding boxes). \\(b) The figure shows the heatmap of all the bounding box positions.} 
    \label{fig3}
\end{figure}
\end{comment}

\label{sec:training}

\subsection{Results}
\label{sec:results}
\begin{table}[ht!]
    \caption{Performance comparison between the proposed approach, FedAvg and FedBN, for different combinations of number of rounds and epochs, on the target metrics. Standard deviations refer to inter-node variability.}
    \label{tab1}
    \centering
\resizebox{16.5cm}{!}{
    \begin{tabular}{lllcc|cc|cc|c}
        \toprule

        \multirow{2}{*}{\textbf{Rounds\phantom{.}}}               & \multirow{2}{*}{\textbf{Epochs\phantom{.}}} & \multirow{2}{*}{\textbf{Methods}} & \multicolumn{2}{c}{\textbf{PPV}}     & \multicolumn{2}{c}{\textbf{TPR}}     & \multicolumn{2}{c}{\textbf{IoU}}     & \multirow{2}{*}{\textbf{\textbf{\phantom{-.}ACC\phantom{-.}}}}\\
        \textbf{}                                      &     &            & \textit{\phantom{.}low\phantom{.}}                    & \textit{\phantom{.}high\phantom{.}}                    & \textit{\phantom{.}low\phantom{.}}                   & \textit{\phantom{.}high\phantom{.}}                  & \textit{\phantom{.}low\phantom{.}}                    & \textit{\phantom{.}high\phantom{.}}                   &\\
        \midrule
        \textbf{---} & 200  & \emph{joint training}   & 69.8 & 61.9      & 80.2 & 48.1      & 52.2                    & 50.8                    & 67.3\\
        \textbf{---} & 200  & \emph{standalone}       & 64.8 & 58.1      & 68.7 & 43.2      & \result{43.9}{2.3}      & \result{42.7}{3.8}      & \result{58.5}{3.2}\\
        \midrule
        \multirow{9}{*}{25} & \multirow{3}{*}{5}   
           & FedAvg~\cite{mcmahan2017communication}   & 55.4 & 29.5      & 78.0 & 25.9      & \result{53.1}{4.1}      & \result{42.0}{11.1}      & \result{72.2}{4.4}\\
        &  & FedBN~\cite{li2021fedbn}                 & 57.5 & 27.9      & 73.3 & 34.1      & \result{51.8}{5.2}      & \result{54.3}{1.4}       & \result{71.8}{4.1}\\
        &  & \textbf{Ours}                            & 57.9 & 33.5      & 73.4 & 33.9      & \result{51.2}{1.8}      & \result{51.3}{6.8}       & \result{70.9}{3.5}\\
        \cmidrule{2-10}     & \multirow{3}{*}{10}   
        & FedAvg~\cite{mcmahan2017communication}      & 55.2 & 28.1      & 77.3 & 25.9      & \result{51.0}{2.6}      & \result{40.8}{11.2}      & \result{71.8}{4.6}\\
        &  & FedBN~\cite{li2021fedbn}                 & 59.1 & 47.1      & 80.5 & 25.3      & \result{54.5}{3.2}      & \result{52.1}{3.8}       & \result{69.0}{1.9}\\
        &  & \textbf{Ours}                            & 62.0 & 47.9      & 77.8 & 31.2      & \result{53.3}{4.0}      & \result{37.3}{9.4}       & \result{70.2}{3.0}\\
        \cmidrule{2-10}     & \multirow{3}{*}{20}
           & FedAvg~\cite{mcmahan2017communication}   & 57.8 & 31.7      & 77.3 & 32.9      & \result{51.4}{2.9}      & \result{44.6}{8.5}       & \result{74.3}{4.5}\\
        &  & FedBN~\cite{li2021fedbn}                 & 58.3 & 29.3      & 74.1 & 34.4      & \result{55.7}{3.9}      & \result{52.1}{3.6}       & \result{72.4}{4.0}\\
        &  & \textbf{Ours}                            & 59.8 & 38.7      & 68.1 & 41.1      & \result{51.1}{3.4}      & \result{40.2}{5.2}       & \result{71.0}{3.4}\\
        
        \midrule

        \multirow{9}{*}{50} & \multirow{3}{*}{5}   
           & FedAvg~\cite{mcmahan2017communication}   & 58.3 & 28.9      & 74.0 & 35.3      & \result{53.2}{4.3}      & \result{54.9}{2.9}       & \result{73.1}{4.3}\\
        &  & FedBN~\cite{li2021fedbn}                 & 63.0 & 73.3      & 85.2 & 32.8      & \result{52.8}{3.4}      & \result{54.0}{1.3}       & \result{71.1}{2.3}\\
        &  & \textbf{Ours}                            & 64.4 & 66.3      & 78.9 & 39.8      & \result{46.2}{4.2}      & \result{44.2}{5.2}       & \result{69.4}{1.6}\\
        \cmidrule{2-10}     & \multirow{3}{*}{10}   
        & FedAvg~\cite{mcmahan2017communication}      & 62.0 & 56.1      & 83.4 & 28.3      & \result{55.1}{4.3}      & \result{47.5}{10.2}      & \result{69.2}{2.0}\\
        &  & FedBN~\cite{li2021fedbn}                 & 63.5 & 72.2      & 94.9 & 30.4      & \result{55.5}{3.7}      & \result{54.3}{4.4}       & \result{69.5}{3.2}\\
        &  & \textbf{Ours}                            & 66.9 & 66.8      & 81.1 & 46.5      & \result{51.2}{2.9}      & \result{46.9}{6.9}       & \result{73.6}{3.8}\\
        \cmidrule{2-10}     & \multirow{3}{*}{20}
           & FedAvg~\cite{mcmahan2017communication}   & 59.3 & 49.8      & 74.1 & 37.3      & \result{56.6}{3.1}      & \result{52.8}{2.1}       & \result{73.9}{4.1}\\
        &  & FedBN~\cite{li2021fedbn}                 & 61.2 & 47.6      & 85.4 & 27.8      & \result{58.5}{5.4}      & \result{48.3}{4.8}       & \result{66.0}{1.5}\\
        &  & \textbf{Ours}                            & 67.6 & 49.8      & 69.5 & 47.9      & \result{50.3}{3.5}      & \result{50.0}{2.9}       & \result{70.4}{3.7}\\
        
        \bottomrule
    \end{tabular}
}
\end{table}

In order to evaluate the performance of our method, FeDETR, we compare its performance to baseline approaches (FedAvg~\cite{mcmahan2017communication} and FedBN~\cite{li2021fedbn}) on standard object detection metrics, namely, accuracy (ACC), precision (PPV), recall (TPR) and intersection over union (IoU). Results are presented in Tab.~\ref{tab1}.

The federated training approach consistently achieves significant accuracy gains over baseline DETR models that utilize a combined dataset from all nodes (\emph{joint}), and even more compared to local training alone (\emph{standalone}). This outcome underscores that training on smaller individual node datasets outperforms a unified dataset in a single node. Furthermore, when comparing federated training on individual node datasets to training on a single node's dataset, there is a notable 12-point increase in accuracy. This suggests that federated training on node-specific data surpasses training on a single consolidated dataset, possibly due to large data distribution shifts between nodes.

The comparison between the three federated approaches, i.e. FeDETR, FedAvg~\cite{mcmahan2017communication} and FedBN~\cite{li2021fedbn}, shows that all of them achieve similar performance in both classification accuracy and object detection metrics. In particular, FeDETR shows generally improved performance at a lower number of rounds, highlighting better convergence properties. Also, while FeDETR seems to achieve better scores in precision rather than recall, we can observe a general trend of improvement on mild/high--severity stenosis, which can be a critical factor for diagnostic purposes. On the other hand, FeDETR shows worse performance on bounding box prediction, based on the IoU metric.

In particular, FeDETR employs a model that shares only 15\% of the total model's parameters, specifically the backbone, omitting the transformer component. This strategic parameter sharing maintains privacy concerns (since weight sharing can be subject to malicious attacks aimed at reconstructing model inputs~\cite{geiping2020inverting,zhu2019deep}) while still delivering competitive performance.

These results collectively underscore the effectiveness of federated training in enhancing accuracy and performance across various scenarios, showcasing its potential for collaborative and privacy-preserving machine learning.

\section{Conclusion}
In this paper, we propose a novel federated learning approach for the assessment of coronary stenosis. Leveraging the Detection Transformer (DETR) model~\cite{carion2020end}, the proposed method includes \emph{fractional flow reserve} (FFR) and \emph{instantaneous wave-free ratio} (iFR) values to accurately evaluate stenosis severity.

In our approach, each node in the federation trains a DETR model to detect high severity stenoses in key frames extracted from angiography videos. At the end of each round, the central server receives the backbones of the models from the nodes and aggregates them.

By taking advantage of federated learning, we address the challenges of privacy concerns and data sharing in the medical domain, ensuring the security and integrity of patient data. Moreover, we provide cardiologists with a reliable tool for accurate assessment of coronary stenosis, contributing to improved patient care and informed medical decision-making.
Our evaluation goes beyond traditional benchmarks by using real data from a dataset collected from five hospitals: we demonstrate that the proposed model is able to achieve competitive performance to state-of-the-art approaches, and even yields better results in the identification of mild/high--severity stenoses, thus showing the robustness and practical applicability of our approach in real clinical settings.

To the best of our knowledge, this work stands as a pioneering endeavor in employing a federated detection transformer, holding applicability within computer vision for detection tasks that hinge upon datasets comprised of non-i.i.d. sub-datasets.
In the future, we aim at improving our method by automating the procedure for the selection of the key frames from angiography videos. Furthermore, we envisage investigating broader applications, such as using our approach for medical examinations that use contrast agents to improve the quality of imaging.

%\section*{Acknowledgments}
%Raffaele Mineo and Amelia Sorrenti are PhD students enrolled in the National PhD in Artificial Intelligence, cycle XXXVII and XXXVIII respectively, course on Health and life sciences, organized by University Campus Bio-Medico of Rome.

\bibliographystyle{unsrtnat}
\bibliography{references}

\end{document}